# Magnetic reconnection between a solar filament and nearby coronal loops

Leping Li[1*], Jun Zhang[1], Hardi Peter[2], Eric Priest[3], Huadong Chen[1], Lijia Guo[2], Feng Chen[2], and Duncan Mackay[3]

[1] Key Laboratory of Solar Activity, National Astronomical Observatories, Chinese Academy of Sciences, Beijing 100012, China
[2] Max-Planck Institute for Solar System Research (MPS), Göttingen 37077, Germany
[3] School of Mathematics and Statistics, University of St. Andrews, St. Andrews, Fife, KY16 9SS, UK

**Magnetic reconnection, the rearrangement of magnetic field topology, is a fundamental physical process in magnetized plasma systems all over the universe[1,2]. Its process is difficult to be directly observed. Coronal structures, such as coronal loops and filament spines, often sketch the magnetic field geometry and its changes in the solar corona[3]. Here we show a highly suggestive observation of magnetic reconnection between an erupting solar filament and its nearby coronal loops, resulting in changes in connection of the filament. X-type structures form when the erupting filament encounters the loops. The filament becomes straight, and bright current sheets form at the interfaces with the loops. Many plasmoids appear in these current sheets and propagate bi-directionally. The filament disconnects from the current sheets, which gradually disperse and disappear, reconnects to the loops, and becomes redirected to the loop footpoints. This evolution of the filament and the loops suggests successive magnetic reconnection predicted by theories[1] but rarely detected with such clarity in observations. Our results on the formation, evolution, and disappearance of current sheets, confirm three-dimensional magnetic reconnection theory and have implications for the evolution of dissipation regions and the release of magnetic energy for reconnection in many magnetized plasma systems.**

Magnetic reconnection[1,2] is considered to play an essential role in the rapid release of the magnetic energy and its conversion to other forms (thermal, kinetic and particle) in magnetized plasma systems (such as accretion disks, solar and stellar coronae, planetary magnetospheres, and laboratory plasmas) throughout the universe. It shows the reconfiguration of the magnetic field geometry. In solar physics, numerous theoretical studies of the magnetic reconnection have been undertaken to explain flares[5], filament eruptions[6], et al. In two dimensional (2D) models, reconnection occurs at an X-point where anti-parallel magnetic field lines converge and reconnect[1,5,6]. So far, many observations of magnetic reconnection signatures, e.g., cusp-shaped post-flare loops[7], loop-top hard X-ray source[3,8], reconnection inflows[3,9] and outflows[3,10,11], flare supra-arcades downflows[12,13], current sheets, and plasmoid ejections[11], have been reported by using remote sensing data. However, to directly observe the details of magnetic reconnection process is difficult, because of the small spatial scale and the fast temporal evolution of the process.

A solar filament is a relatively cool and dense plasma structure in the corona suspended above a magnetic polarity inversion line (PIL), with ends rooted in regions with opposite magnetic polarity. Its spine, a narrow ribbon-like structure through the full filament, consists of horizontal and parallel threads[14] when viewed from above. In the region around a filament, the plasma-beta, i.e., the ratio of thermal to magnetic energy density, is below unity. Because of the high electric conductivity, the magnetic flux is frozen into the coronal plasma[1,3]. The magnetic field topology is therefore often outlined by plasma trapped in the coronal magnetic structures, (such as filaments[14,15] and coronal loops[3]). The change of these coronal structures hence suggests a change of magnetic topology[3]. Erupting filaments usually reconnect with ambient coronal structures, such as a coronal hole[16], another filament[17], or active region loops[18]. However, the present observations give much more detail of the reconnection process.

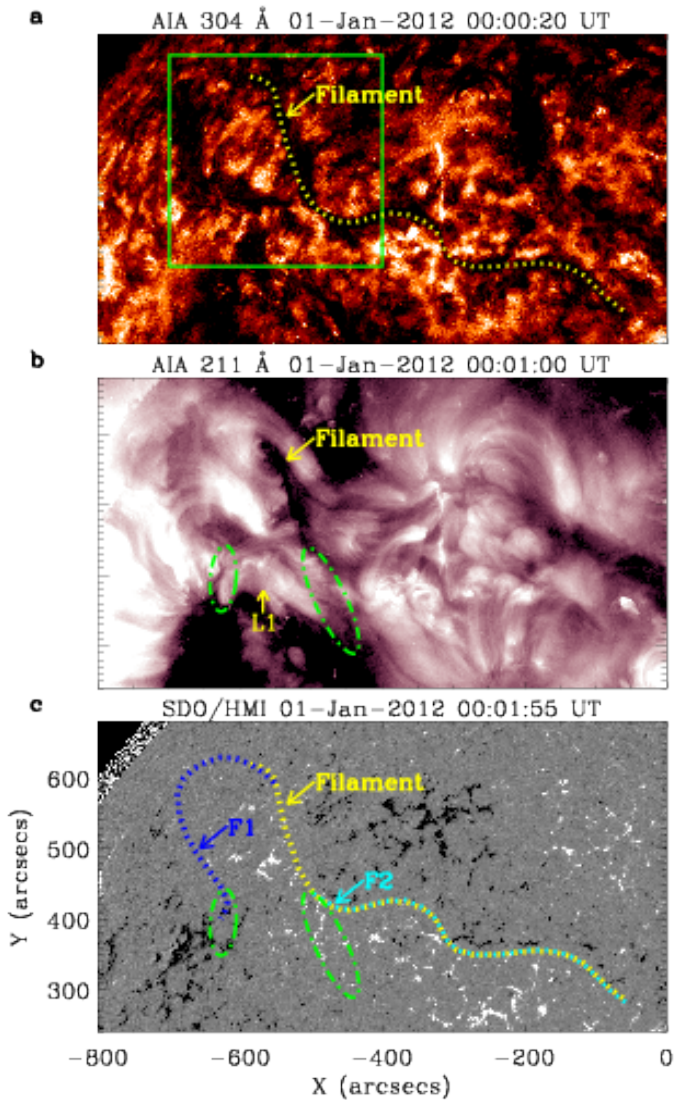

**Figure 1 | SDO observations of a quiet Sun filament and the context coronal loops and magnetic fields.** Panels (a-c) separately show AIA 304 Å (a) and 211 Å (b) images, and an HMI line-of-sight (LOS) magnetogram (c). The green rectangle in (a) denotes the field of view (FOV) in Figs. 2a-2c. The ellipses in (b-c) enclose the footpoints of the loops L1. The yellow, blue, and cyan dotted lines in (a) and (c) separately indicate the original filament and two newly reconnected filaments F1 and F2.

The Atmospheric Imaging Assembly[19] (AIA) and Helioseismic and Magnetic Imager[20] (HMI) onboard the Solar Dynamics Observatory[21] (SDO) provide full-disk multiwavelength images of the solar atmosphere and light-of-sight (LOS) magnetograms, with time cadences and spatial sampling of 12 s and 45 s, and 0.6"/pixel and 0.5"/pixel, respectively. On January 1, 2012, a quiet Sun filament, with a length of about 720" and width of about 13", was observed at around 00:00:20 UT by AIA 304 Å (~0.05 MK) in the northern hemisphere (see Fig. 1a). It was located above a PIL between opposite polarity magnetic field concentrations (see the yellow dotted line in Fig. 1c). Its northeastern endpoint roots in positive magnetic fields, while the southwestern one in negative fields.

The northeastern part of the filament is clearly identified in simultaneous AIA higher temperature channels, e.g., 335 Å (~2.5 MK) and 211 Å (~1.9 MK), as an elongated dark structure, but the southwestern part is overlaid by dense closed loops (see Fig. 1b). To the southeast of the filament, a set of hot coronal loops L1 is present (see Fig. 1b), rooted in opposite magnetic polarities (see the green ellipses in Figs. 1b-1c). This filament is also observed by the Extreme UltraViolet (EUV) Imager (EUVI) onboard the Solar Terrestrial Relation Observatory[22] B (STEREO-B) (see Supplementary Fig. S1). It is located near the northwestern solar limb (see Supplementary Fig. S2) in the field of view (FOV) of EUVI, which supplies successive 304 Å (~0.05 MK) and 195 Å (~1.5 MK) images with spatial sampling and time cadences of 1.6"/pixel and 1.6"/pixel, and 10 minutes and 5 minutes, respectively.

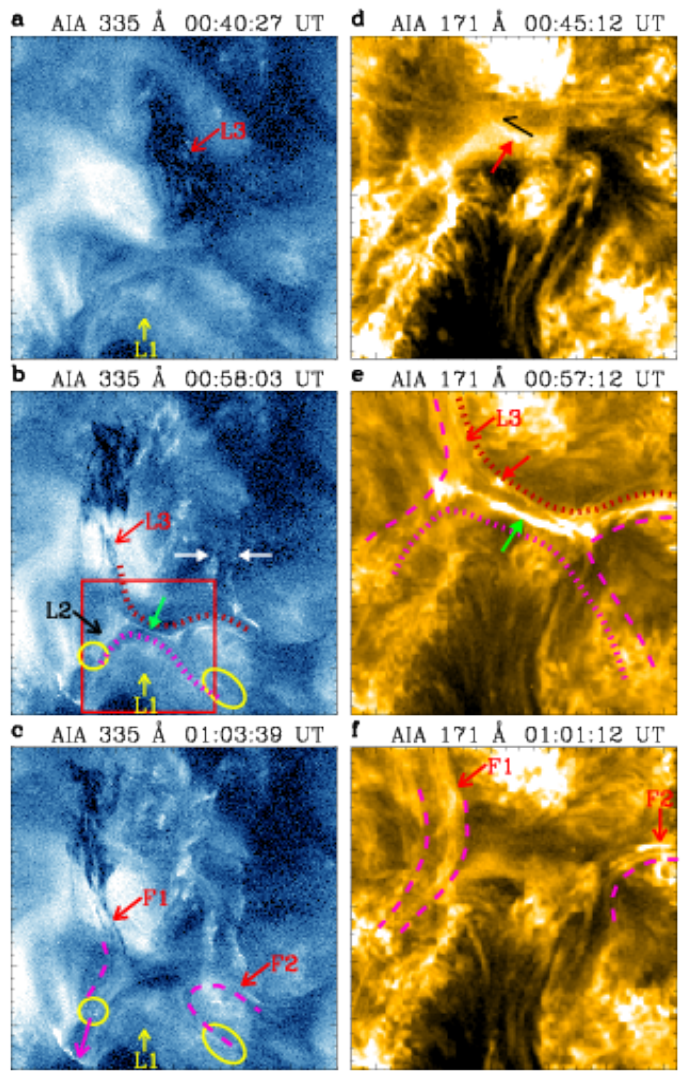

**Figure 2 | Magnetic reconnection between the filament and nearby coronal loops.** Panels (a-f) show AIA 335 Å (a-c) and 171 Å (d-f) images. The red rectangle in (b) shows the FOV in Figs. 2d-2f and Fig. 4. The red and pink dotted lines in (b) and (e) indicate the filament threads L3 and the loops L2, and the pink dashed lines in (c) and (e-f) F1 and F2. The yellow circles and ellipses in (b-c) enclose the footpoints of L2, F1, and F2. The

green solid arrows in (b) and (e) mark the current sheets, two white arrows in (b) the flare ribbons, the pink and black arrows in (c-d) the propagations of the southern footpoints of F1 and a plasmoid, and the red solid arrow in (d-e) the plasmoids.

Over the course of a little more than an hour, the whole structure of the filament underwent dramatic changes due to magnetic reconnection (see the blue and cyan dotted lines in Fig. 1c). These can be seen in the image sequence of Fig. 2, which shows the evolution of ~2.5 MK (335 Å; blue colored) and ~0.9 MK (171 Å; orange colored) plasma. How does the magnetic reconnection happen? From about 00:10:00 UT, the northeastern part of the filament slowly rises eastward with a mean speed of 4.8 km s$^{-1}$. It then quickly erupts with a speed of 43.3 km s$^{-1}$ and an acceleration of 10.7 m s$^{-2}$, and is dispersed into many parallel dark threads L3 along the spine when viewed from above (see Fig. 2a and Supplementary Fig. S3). Two flare ribbons (white arrows in Fig. 2b; see also Supplementary Movies 1 and 4) appear underneath the erupting filament, supporting the classical flare models[5,6]. The southwestern part of the filament does not erupt due to the constraint of the overlying coronal loops (see Fig. 1b). The eruption of the northeastern part of the filament is also observed by STEREO-B/EUVI images simultaneously (see Supplementary Fig. S2 and Movie 2).

During the eruption, the filament L3 encounters nearby coronal loops L2 and interact with them, forming an X-type configuration with sheet-type and cusp-type structures at the interface (see the red and pink dotted lines in Fig. 2b and Supplementary Fig. S4). This X-type configuration is suggestive of magnetic reconnection[3]. The sheet-type structures in Fig. 2b (green arrow) indicate the positions of current sheets[1-6], which are the dissipation regions of magnetic energy during the reconnection process. The X-type loop structure is mainly observed in the channels of AIA that show hotter plasma, e.g., 335 Å and 211 Å. However, only the filament part of the structure is visible in the cooler channels of AIA, such as 193 Å (~1.5 MK), 171 Å (~0.9 MK), 131 Å (~0.6 MK), or 304 Å (~0.05 MK). Brightenings at the loop footpoints (see the yellow circle and ellipse in Fig. 2b) are noticeable during the reconnection process (see Supplementary Fig. S4). They are heated by thermal conduction and/or non-thermal particles that are produced in the reconnection region[1,2,4-6] and propagate along the newly reconnected magnetic field lines. The current sheets gradually diffuse and disappear. The northeastern part of the filament then reconnects to the eastern leg of the loops L2, forming a new filament F1, while the southwestern part reconnects to the western leg of L2, forming another new filament F2 (see Figs. 2c and 1c). The southern footpoints of F1 show an apparent motion toward the southeast with a speed of 72 km s$^{-1}$ (see pink arrow in Fig. 2c and Supplementary Fig. S5), indicating a continuing reconnection process.

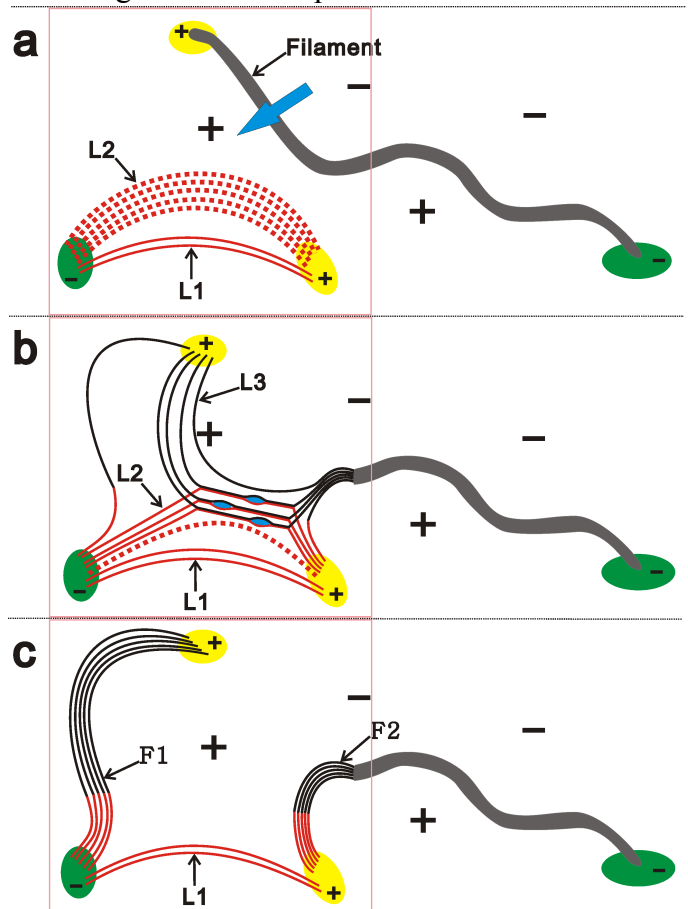

**Figure 3 | Schematic diagrams of the magnetic reconnection.** Panels (a-c) show the magnetic field topology before (a), during (b), and after (c) the magnetic reconnection. The yellow (plus) and green (minus) ellipses (signs) show the positive and negative magnetic fields. The grey thick lines indicate the filaments. The red, black, and red-black lines denote the magnetic field lines. The cyan arrow in (a) shows the erupting direction of the filament. The cyan ellipses in (b) indicate the plasmoids. L1 and L2 mark the loops, L3 the initial filament threads, and F1 and F2 the newly reconnected filaments. The pink rectangles in (a-c) indicate the FOV in Figs. 2a-2c.

The southern (eastern) footpoints of F1 (F2) are the same as the eastern (western) footpoints of L2, see the yellow circles (ellipses) in Figs. 2b-2c. This clearly indicates an interchange magnetic reconnection process between the magnetic field lines of the filament and the coronal loops.

The magnetic flux in the southern footpoint region of F1, ~9.5×10$^{19}$ Mx (see Supplementary Fig. S5), has to equal the flux at the western footpoint of the loops L2 before the reconnection and of the northern footpoint of F1 after the reconnection (see Figs. 2b-2c). Thus, twice the flux of the southern footpoint, 1.9×10$^{20}$ Mx, has been reconnected.

The reconnection of the filament and the apparent motion of the southern footpoints of F1, are also observed by EUVI 304 Å and 195 Å images (see Supplementary Fig. S2 and Movie 2). The filament material of the eastern part of F2 flows westward into the unerupted filament part, and that of F1 flows down into its two footpoints. Both of the structures become invisible eventually in the cool lines (see Supplementary Movies 1-2 and 4) since there is no cool filament material in them, suggesting that the newly reconnected magnetic field structures of both F1 and F2 cannot support dense filament material.

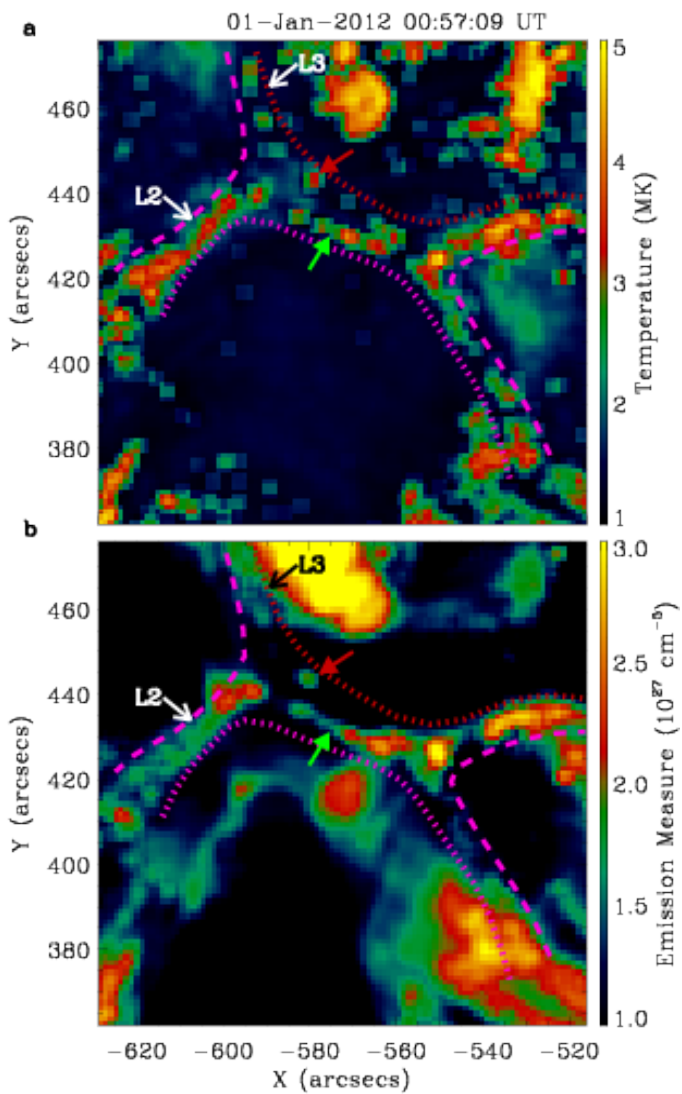

**Figure 4 | Maps of temperature and emission measure (EM).** Panels (a-b) show the temperature (a) and EM (b) maps obtained using the differential EM (DEM) analysis method[28]. Similar to Fig. 2e, the dotted lines show the filament threads L3 and the loops L2, and the dashed lines the newly reconnected filaments. The green and red arrows mark a current sheet with plasmoids.

While X-type structures indicative of magnetic reconnection have been reported before[3], here we can also investigate more details of the formation and evolution of plasmoids and current sheets. For this we refer also to Fig. 3 which gives an overall cartoon-picture of what is happening in Fig. 2.

When the erupting filament L3 meets the neighbouring loops L2, a bright plasmoid appears and propagates northeastward (red and black arrows in Fig. 2d). Also, a straight current sheet gradually brightens at the interface and later diminishes (see Supplementary Fig. S6). Subsequently, more filament threads reconnect with the preexisting loops and form more current sheets with propagating bright plasmoids (red and green solid arrows in Fig. 2e; see also Supplementary Figs. S3 and S6-S7). After the current sheet disappears, the filament L3 connects all the way down to the southern footpoint of F1 (see Fig. 2f). In this process they move away from the reconnection region with speeds of about 40 km s$^{-1}$ to 170 km s$^{-1}$ (see Supplementary Fig. S6).

Investigating the temporal evolution, at least twelve individual current sheets are identified, moving southeastward with velocities of about 35 km s$^{-1}$ to 120 km s$^{-1}$ (see Supplementary Fig. S6). Some of these current sheets merge (see Supplementary Figs. S3 and S6). They last for about 15 minutes, and have lengths of about 25" to 41", and widths of about 1.0" to 1.6" (see Supplementary Fig. S7).

Furthermore, many bright plasmoids appear within the current sheets and propagate bi-directionally along them, and then further along the filament or the loops (see Supplementary Fig. S6). They move with speeds of 100 km s$^{-1}$ to 300 km s$^{-1}$, have a width of about 0.8" to 1.7", and are slightly ellipsoidal (see Supplementary Fig. S7). The plasmoids tend to move towards the erupting filament as they mostly move northeastward (see Supplementary Movie 3).

Details of the plasmoids in the reconnecting current sheets are clearly revealed by maps of temperature and emission measure (EM). These are obtained using a differential EM (DEM) analysis[23] of six AIA EUV (94 Å, 335 Å, 211 Å, 193 Å, 171 Å, and 131 Å) channels (see Fig. 4 and Supplementary Movie 3). Consistent with the EUV observations, X-type structures, current sheets, and plasmoids are also identified in these temperature and EM maps shown in Fig. 4. The maps are cotemporal with the AIA 171 Å channel displayed in Fig. 2e.

Inside the currents sheets (i.e., the bright linear structures in Fig. 2e indicated by a green arrow), we see chains of enhanced temperature and EM (Fig. 4). These are plasmoids, one of which is pointed out by the solid red arrows in Fig. 4 and in Fig. 2e. In the plasmoids, the temperature ranges from 3.7 MK to 6.5 MK. While there is little variation of the emission along the current sheets (Fig. 2e), there is a strong variation in temperature between the hot plasmoids and the cooler regions between them, which have less than half the temperature of the plasmoids (Fig. 4a). Still, the current sheets are significantly hotter than the filament threads measured before they reconnect, due to the heating produced by reconnection.

The same applies for the EM. The density, $n=(EM\ L^{-1})^{0.5}$, of the current sheets can be estimated to range from $3.6\times10^9$ cm$^{-3}$ to $5.3\times10^9$ cm$^{-3}$, with the higher values found in the plasmoids.

The spine of a filament consists of dense plasma concentrated around the centre of a longer magnetic flux tube[14-18], and so the disconnections and reconnections of the filament spine indicate a reconfiguration of magnetic field topology of the filament. Our observations represent a clear example of such reconnection and provide details of the process. In total a magnetic flux of $1.9\times10^{20}$ Mx in the filament and the loops reconnects in about 15 minutes, resulting in a mean reconnection rate of $2.1\times10^{17}$ Mx s$^{-1}$. This is an order of magnitude less than for strong X- or M-flares[24], but is reasonable for reconfiguration in a non-flaring quiet Sun region. In the process there is a successive piles up of magnetic field lines of the filament and the loops (see Fig. 3b), as in the flux-pile-up magnetic reconnection regime[3]. Thereafter, X-type structures and multiple current sheets gradually form (see Fig. 3b). The appearance of bright plasmoids in the current sheets (see Figs. 2e, 3b and 4) suggests the presence of plasmoid instabilities[25] during the reconnection. Magnetic reconnection releases and converts magnetic energy to kinetic energy[1,4,10-13] and thermal energy[3] (as well as fast particle energy). This is consistent with our observations, where the plasmoids are accelerated along the current sheets and show an enhanced temperature. The new details of the reconnection process reported here, such as the evolution of multiple current sheets with many plasmoids, can help us to understand, constrain, advance, and improve three-dimensional magnetic reconnection theory and our understanding of this fundamental process of energy conversion in the cosmos.

**Acknowledgements**

The authors are indebted to the SDO and STEREO teams for providing the data. The work is supported by the National Science Foundations of China (G 11533008, 11221063, 11303050, 11322329), the National Basic Research Program of China under grant G2011CB811403, and the CAS KJCX2-EW-T07.

**Author contributions**

L. P. L. analyzed the data, wrote the text and led the discussion. J. Z., H. P., E. P., H. D. C., L. J. G., F. C., and D. M. contributed to the data interpretation and helped to improve the manuscript.

**Additional information**

Supplementary information is available in the online version of the paper. Reprints and permissions information is available online at www.nature.com/reprints. Correspondence and requests for materials should be addressed to L. P. L.

**Competing financial interests**

The authors declare no competing financial interests.

# Supplementary Material for
# Magnetic reconnection between a solar filament and nearby coronal loops

**Leping Li**[1*], **Jun Zhang**[1], **Hardi Peter**[2], **Eric Priest**[3], **Huadong Chen**[1], **Lijia Guo**[2], **Feng Chen**[2], and **Duncan Mackay**[3]

[1]Key Laboratory of Solar Activity, National Astronomical Observatories, Chinese Academy of Sciences, Beijing 100012, China

[2] Max-Planck Institute for Solar System Research (MPS), Göttingen 37077, Germany

[3] School of Mathematics and Statistics, University of St. Andrews, St. Andrews, Fife, KY16 9SS, UK

Figures S1-S7
Movies M1-M4

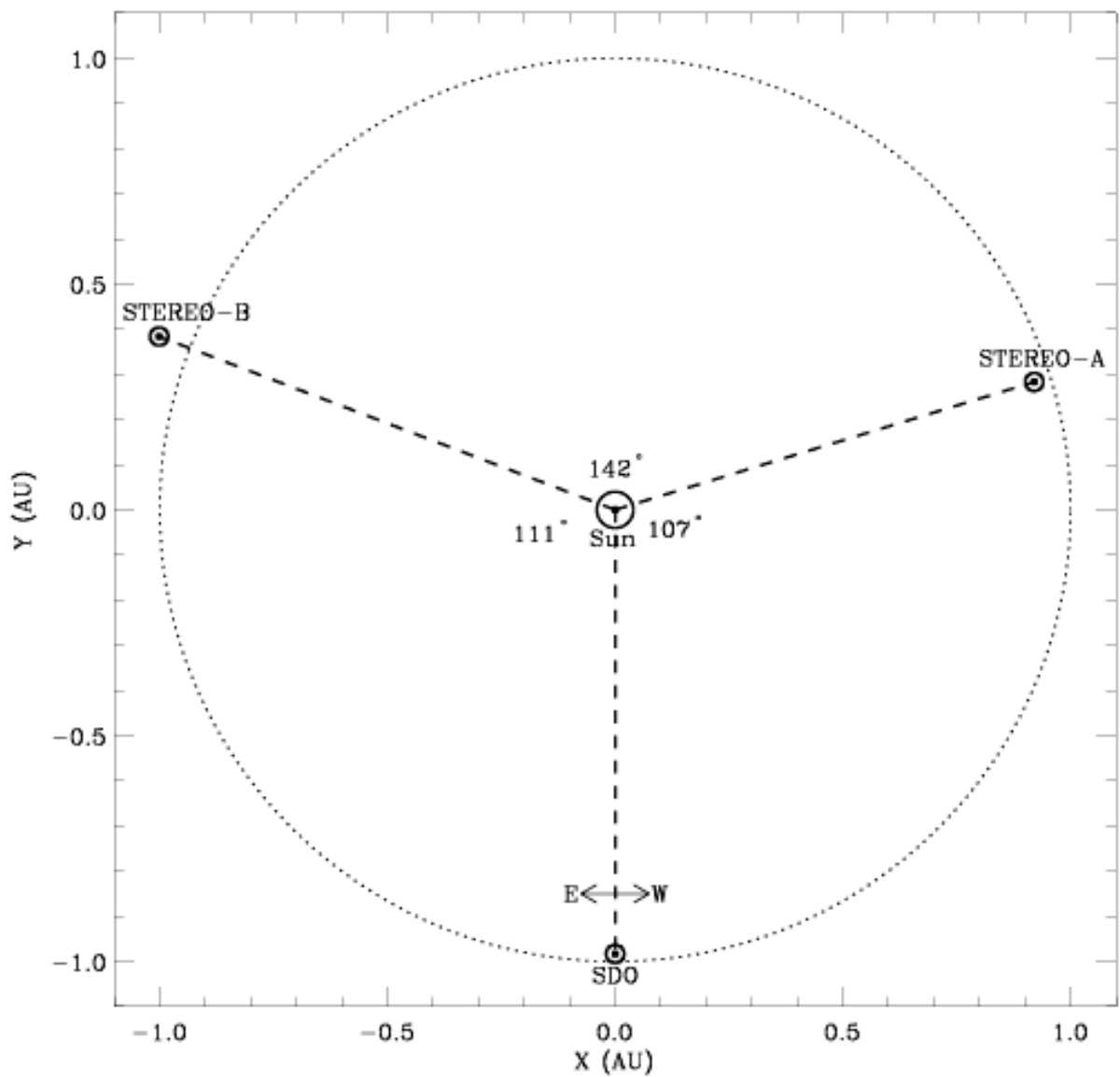

**Figure S1. Positions of the SDO and STEREO A and B satellites at 01:00:00 UT on January 1, 2012.** The dotted circle represents the Earth orbit at 1 AU. The angles between these three satellites are denoted by the numbers in the plot. E and W show the directions, i.e., east and west, in the field of view (FOV) of SDO. As the filament is located near the northeastern limb in the FOV of SDO/AIA images, it can also be observed by STEREO-B, rather than STEREO-A.

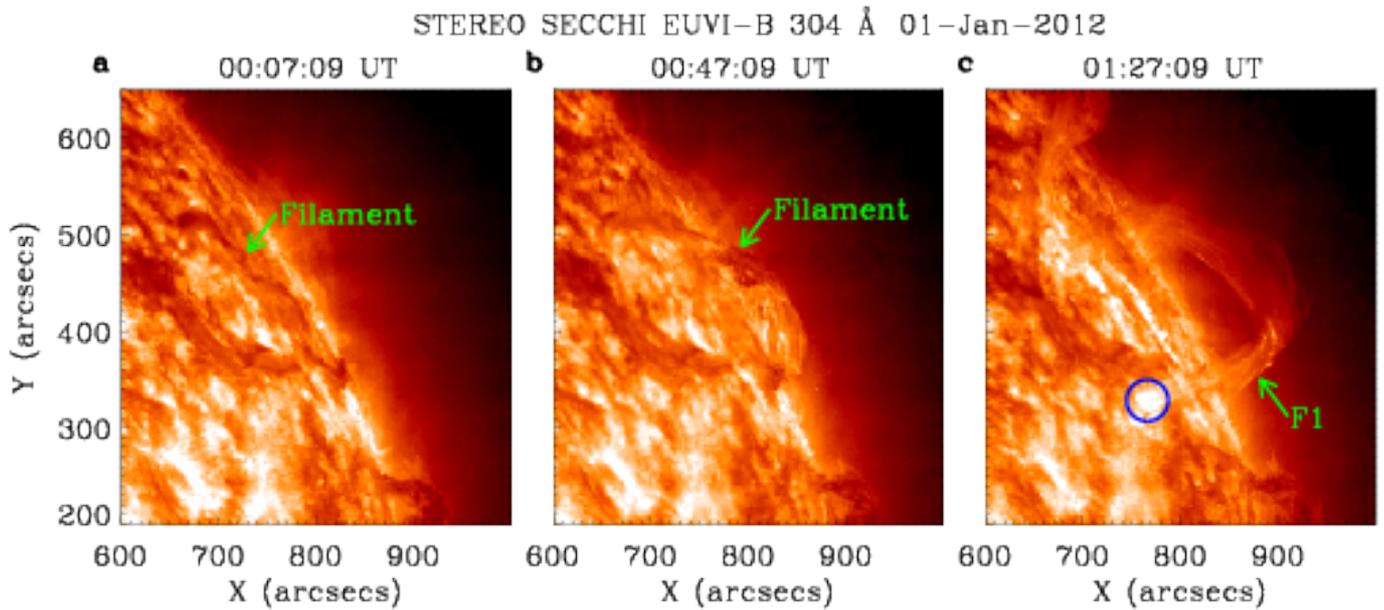

**Figure S2. Evolution of the filament observed by STEREO-B.** Panels (a-c) separately show EUVI 304 Å images before (a), during (b), and after (c) the magnetic reconnection. A blue circle in (c) encloses the southern footpoints of the newly reconnected filament F1. This circle is pointing at roughly the same location as the circle at the footpoint of F1 in Fig. 2c. Because of the separation angle of 111º between STEREO-B and SDO, this review is in approximate quadrature.. The eruption of the filament can be followed through this image sequence. In particular panel (c) shows the filament after the reconnection, now connects to a new footpoint highlighted by the blue circle.

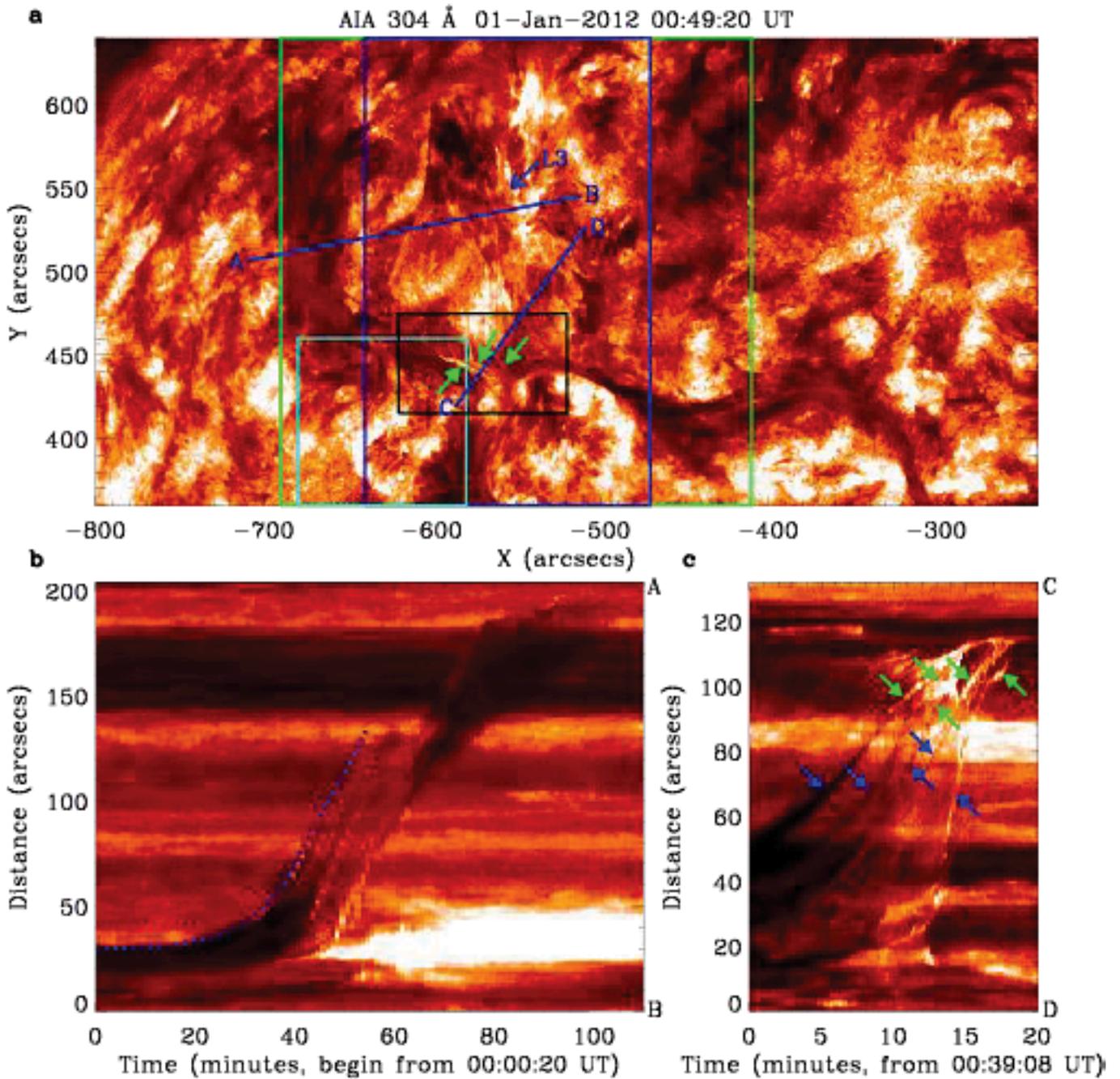

**Figure S3. Filament eruption observed by SDO/ AIA.** Panel (a) displays AIA 304 Å image at 00:49:20 UT. Panels (b-c) display time-space plots of a series of AIA 304 Å images, respectively for the time evolution along lines AB (b) and CD (c) marked in (a). The green arrows in (a) and (c) mark the current sheets, a dotted line in (b) the filament eruption, and the blue arrows in (c) the filament threads. Panel (b) shows that the filament rises slowly at the beginning, and then accelerates and erupts. During the eruption, the filament is dispersed into threads when viewed from above, marked by blue arrows in (c). When these threads encounter the loops, they became bright current sheets at the interfaces, denoted by green arrows in (a) and (c). Due to the filament eruption, the current sheets move toward the southeast (towards lower left in panel a), and some of them merge (c; see also Fig. S6d). For comparison the green rectangle shows the FOV of Figs. 2(a-c), the blue rectangle the FOV of Fig. S4a, the cyan rectangle the FOV of Figs. S5(a-b), and the black rectangle the FOV of Figs. S6(a-b) and Fig. S7a.

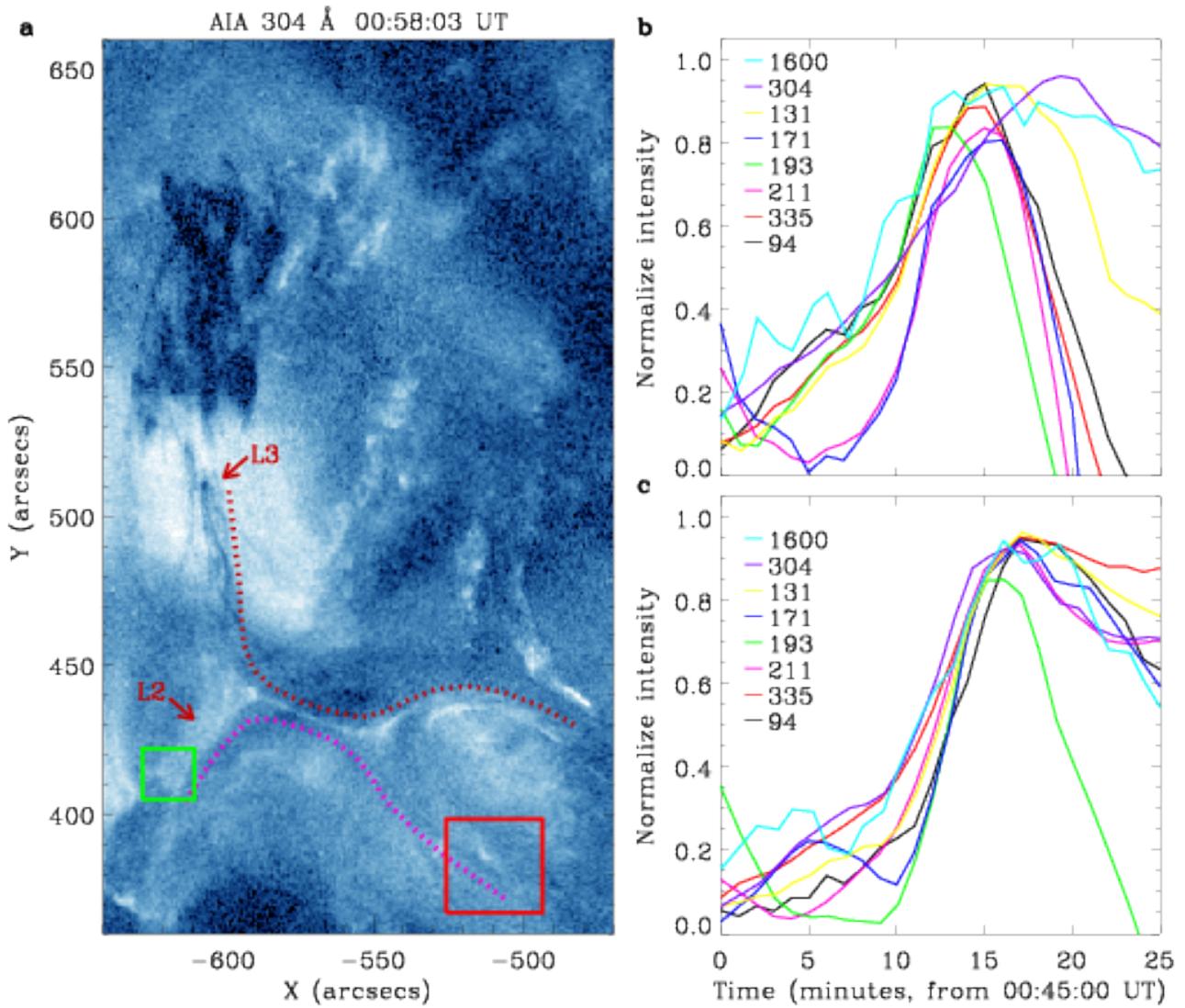

**Figure S4. Brightening at the footpoints of the loops L2 observed by SDO/ AIA.** Panel (a) displays AIA 335 Å image at 00:58:03 UT. Panels (b-c) show the normalized light curves of eight AIA UV channels (wavelengths in Å given with the plots), in the green (b) and red (c) rectangles as displayed in (a). Similar as in Fig 2b the red and pink dotted lines in (a) outline the filament threads L3 and the loops L2, respectively, forming an X-type structure. All the light curves in (b-c) show a clear increase of the emission, simultaneous in both footpoints, representing the brightening at the footpoint regions of L2 over a wide range of temperatures. For comparison, the FOV of (a) is marked as blue rectangle in Fig. S3a.

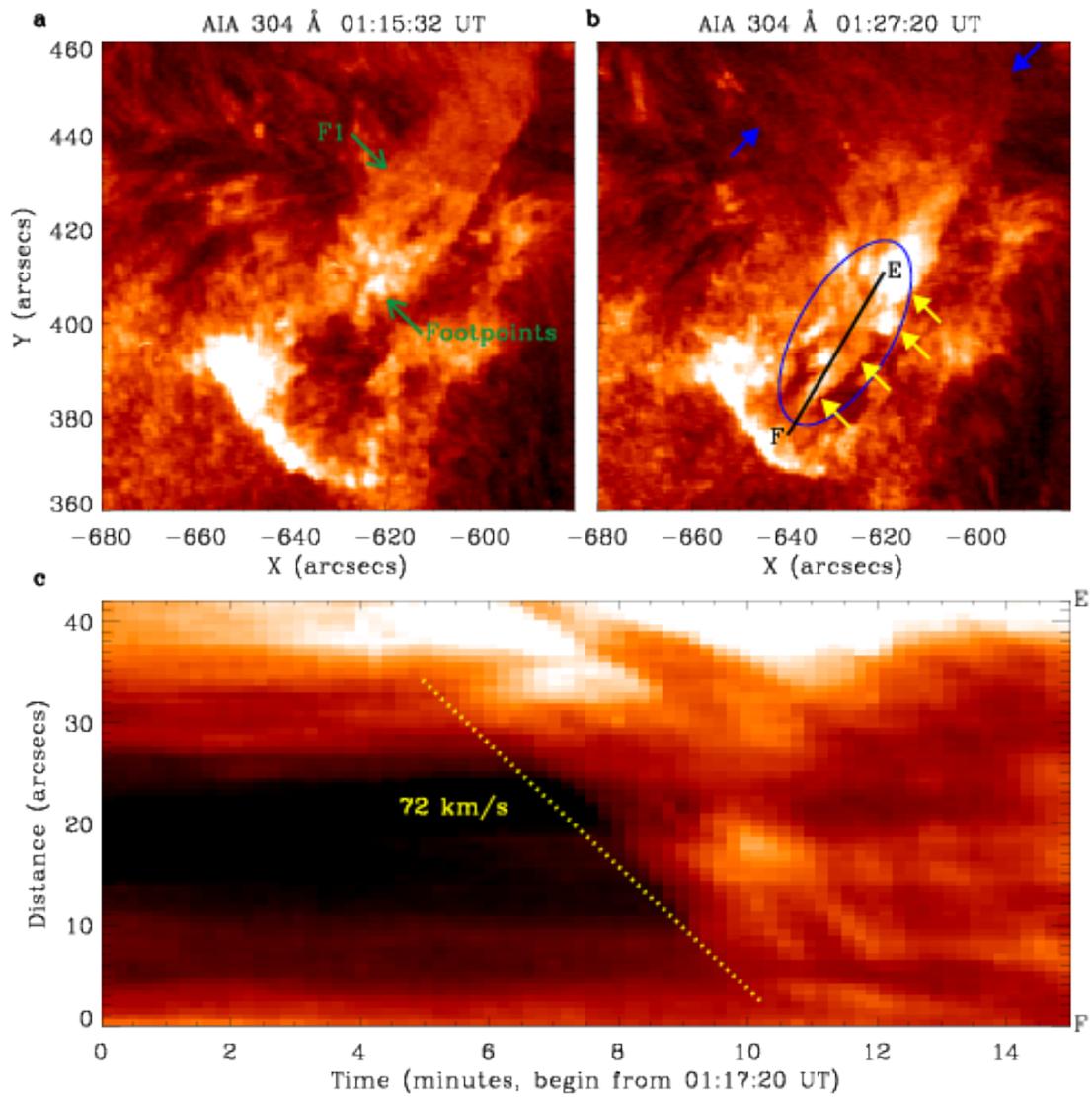

**Figure S5. Brightening propagation of the southern footpoints of F1 observed by SDO/ AIA.** Panels (a-b) show AIA 304 Å images with the FOV as indicated in Fig. S3a (cyan rectangle). Panel (c) displays a time-space plot of a series of AIA 304 Å images along the line EF marked in (b). The yellow and blue arrows in (b) denote the filament footpoints and threads, respectively. The blue ellipse in (b) encloses the footpoint region. The yellow dotted line in (c) indicates the apparent propagation of the brightening at the footpoint with a speed of just above 70 km $s^{-1}$. The magnetic flux of the southern footpoint region of F1 as noted in the main text is measured in the footpoint region of F1 (blue ellipse in b).

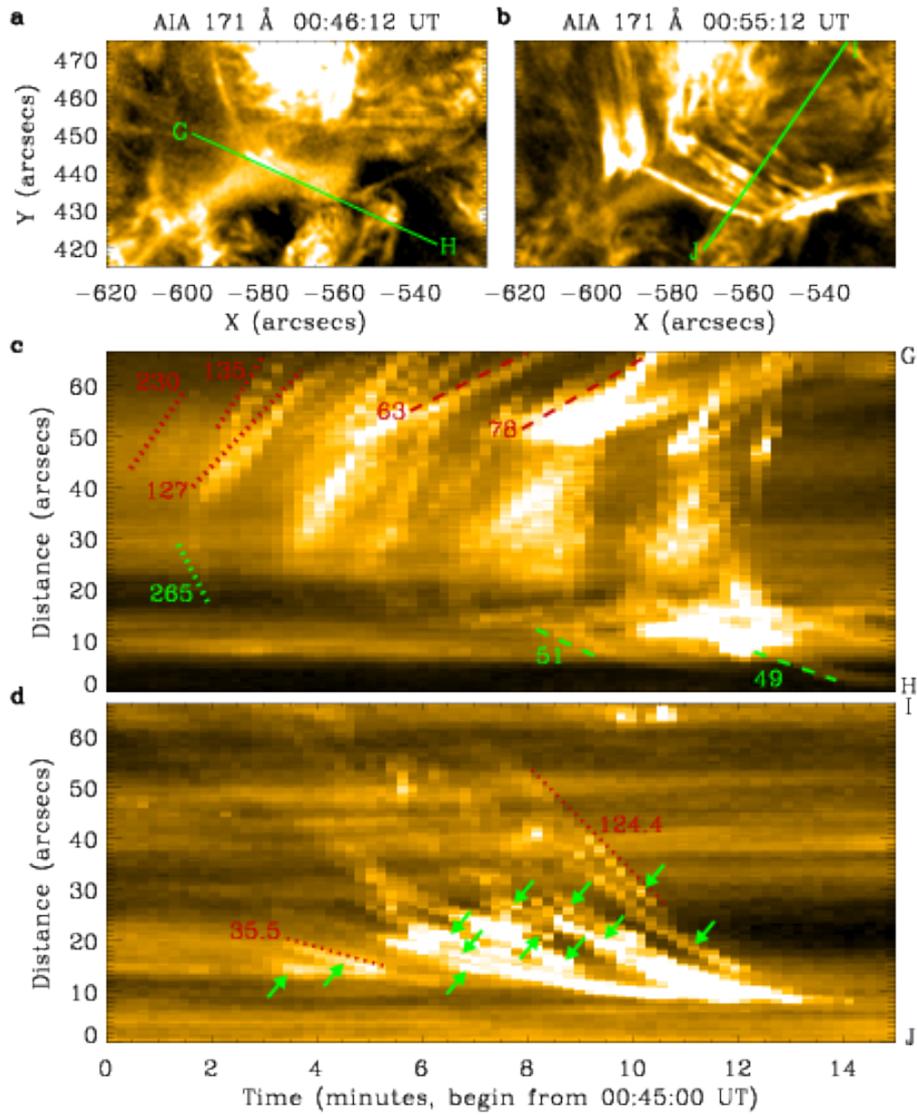

**Figure S6. Evolution of the current sheets and plasmoids observed by SDO/AIA.** Panels (a-b) show AIA 171 Å images (FOV outlined in Fig. S3a as black rectangle). Panels (c-d) display time-space plots of a series of AIA 171 Å images along the lines GH (c) and IJ (d) as marked in (a) and (b), respectively. The dotted and dashed lines in (c) denote the motions of the plasmoids and the newly reconnected filament threads. The green arrows and red dotted lines in (d) show the current sheets and their sideward motions. The respective speeds are denoted by the numbers (km s$^{-1}$) in the plots. Plasmoids mostly move towards northeast, i.e., upper left (b). So in general the plasmoids tend to move toward the erupting filament. In contrast, the current sheets move toward the southeast (lower left) due to the filament eruption (see also Fig. S3c). At least twelve individual current sheets are detected from panel (d). Some of them disappear, and some of them merge. After some 15 minutes they totally disappear at last.

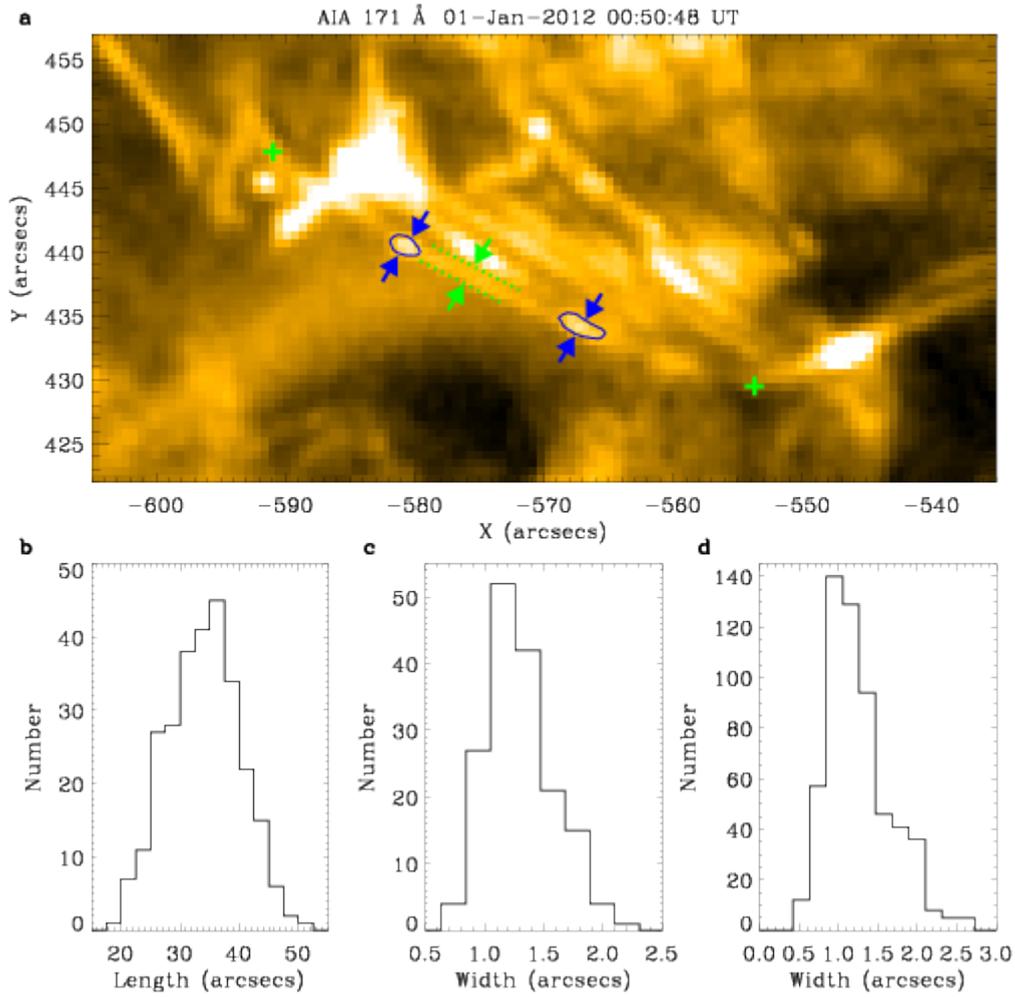

**Figure S7. Measurements of current sheets and plasmoids observed by SDO/AIA.** Panel (a) displays a snapshot AIA 171 Å image at 00:50:48 UT, with the same FOV of Figs. S6(a-b). Panels (b-d) display histograms of properties of current sheets and plasmoids observed during 15 minutes around the reconnection of the filament and the loops. These are the length (b) and the width (c) of current sheets, and the width of plasmoids (d). In (a), the green pluses mark two endpoints of a current sheet, the blue contours and green dotted lines denote the boundaries of two plasmoids and a current sheet, respectively, and the blue and green arrows the widths.

**Movie M1. SDO observations of the temporal evolution of magnetic reconnection between the erupting filament and its nearby coronal loops.** The panels show emission from the hot corona down to the cool chromosphere and the magnetic field on the solar surface according to the following table:

| AIA 94 Å<br>Fe XVIII<br>7.2 MK | AIA 335 Å<br>Fe XVI<br>2.5 MK | AIA 211 Å<br>Fe XIV<br>1.9 MK | |
|---|---|---|---|
| AIA 193 Å<br>Fe XII<br>1.5 MK | AIA 171 Å<br>Fe IX<br>0.9 MK | AIA 131 Å | |
| | | Fe VIII<br>0.6 MK | Fe XXI<br>10.0 MK |
| AIA 304 Å<br>He II<br>0.05 MK | AIA 1600 Å<br>Chromosphere<br>0.01 MK | HMI<br>LOS magnetograms<br>surface | |

The time cadence is 1 minute for AIA images, and 3 minutes for HMI line-of-sight (LOS) magnetograms. The black boxes in all panels show the FOV of Figs. 2(a-c), and the white boxes that of Figs. 2(d-f) of the main text.

This animation can be downloaded from http://ddl.escience.cn/f/xUsG

**Movie M2. STEREO-B SECCHI EUVI observations of the temporal evolution of the eruption and reconnection of the filament.** This shows the erupting filament from the roughly quadrature to AIA observations (see Fig. S1), so the erupting filament is seen from the northwestern side. The two panels show plasma at about 0.05 MK at 304 Å (left) and at 1.5 MK at 195 Å (right) at the same FOV as Fig. S2. The cadence is 10 minutes and 5 minutes, respectively. This animation can be downloaded from http://ddl.escience.cn/f/xUsB

**Movie M3. SDO observations of the temporal evolution of current sheets.** Similar to Movie M1, but for a smaller FOV (white boxes in Movie M1, same as Figs. 2d-f and Fig. 4). In particular the time cadence is higher, now at 12 s. Here we replaced the two bottom right panels by the emission measure (EM) and the temperature (TE) maps. The EM and TE maps show the temporal evolution of Fig. 4 in the main text. This animation can be downloaded from http://ddl.escience.cn/f/xUsE

**Movie M4. Composite multi-color images showing the magnetic reconnection process between the erupting filament and the loops.** The movie shows a composite of AIA images in different wavelength channels (and thus at different plasma temperatures): the red channel shows 335 Å (~2.5 MK), the green 193 Å (~1.5 MK), and the blue 304 Å (~0.05 MK). The FOV is the same as in Movie M1, but here the cadence is faster at 12 s. This animation can be downloaded from http://ddl.escience.cn/f/xUsC